# Academics evaluating academics: a methodology to inform the review process on top of open citations

Federica Bologna
Department of Classical Philology and Italian Studies
University of Bologna, Italy
federica.bologna3@studio.unibo.it

Angelo Di Iorio
Department of Computer Science and Engineering
University of Bologna, Italy
angelo.diiorio@unibo.it

Silvio Peroni
Research Centre for Open Scholarly Metadata
Department of Classical Philology and Italian Studies
University of Bologna, Italy
silvio.peroni@unibo.it

Francesco Poggi
Department of Communication and Economics
University of Modena and Reggio Emilia, Italy
fpoggi@unimore.it

## ABSTRACT

In the past, several works have investigated ways for combining quantitative and qualitative methods in research assessment exercises. In this work, we aim at introducing a methodology to explore whether citation-based metrics, calculated only considering open bibliographic and citation data, can yield insights on how human peer-review of research assessment exercises is conducted. To understand if and what metrics provide relevant information, we propose to use a series of machine learning models to replicate the decisions of the committees of the research assessment exercises.

## 1 Introduction

In the past ten years, an increasing number of works have focused on the divide and connection between qualitative and quantitative science studies [10, 11, 20, 23, 30].

An important field of application of both quantitative and qualitative approaches for research assessment is that shown in the national scientific qualification exercises done worldwide. While they can be structured differently according to the legal landscape and social norms of the research ecosystem of the country they are executed, such exercises aim at deciding whether a scholar can apply to professional academic positions as Associate Professor (AP) and Full Professor (FP), and usually consist of a quantitative and qualitative evaluation process, which first makes use of bibliometrics and standard peer-review processes, where candidates' CVs are assessed by a committee (a.k.a. the commission) to get to the final decision.

In this study, we aim at drawing from some experiments done in the past [3] and expanding this discussion on the use of *open* bibliographic and citation data in the context of research assessment exercises, focussing on the possible relation that exists between quantitative information (provided by data) and qualitative evaluations (provided by humans).

The open availability of bibliographic and citation data, and their use in the study of bibliometric disciplines [5, 16, 21, 31], has been advocated by many scholars in the recent past, often led by international initiatives such as the Initiative for Open Citations (I4OC, https://i4oc.org) and the Initiative for Open Abstracts (I4OA, https://i4oa.org). This push has also resulted in numerous projects, such as COCI [14], OUCI [4], Unpaywall [9], and the extensions of VOSviewer [8] to handle open scholarly metadata and open citations, just to mention a few.

In this paper, we present a *methodology* we aim to use for answering the following research questions:

- Can citation-centric metrics computed using open access datasets provide insights on the human evaluation of academics?
- What role does the relationship between the candidate and the commission play in the reviewing phase of research assessment exercises?

By suggesting the use of bibliographic metadata and citation data from existing open datasets, we devise a set of bibliographic and citation-based metrics using a top-down process and combining popular metrics (e.g. bibliographic coupling) to describe the relationship between the candidates and the commision of research assessment exercises. Using these metrics, we propose to adopt a few machine learning methods to show whether and which citation-based and non-citation-based metrics may inform human decisions within research assessment exercises.

## 2  Methods and materials

This section introduces all the methods and material used for our study.

### 2.1  Citation Network Analysis and Metrics

The main idea of our work is to calculate some metrics on the research production of the candidates participating in research assessment exercises, exclusively from open data, and to study if these metrics could give insights on the peer-review process. The metrics we take into account are organized in two groups. Table 1 shows the first one. It contains some well-established measures to assess the productivity and impact of scholars, thus they do not require further explanation.

| ID | Description |
|---|---|
| cand | overall number of publications authored by the candidate found in the open sources of use |
| books | number of books authored by the candidate |
| articles | number of journal articles authored by the candidate |
| other_pubbs | number of other kinds of publications (e.g. proceedings articles and workshop papers) authored by the candidate |
| co-au | number of publications authored by both the candidate and at least one member of the commission |

Table 1: **The basic metrics extracted from the list of publications of each candidate**

The last metric in Table 1 (i.e. co-au) deserves some discussion. It is meant to investigate if the relation of co-authorship between the candidates and the commission members has played some role in the evaluation process. The idea of studying the co-authorship network between candidates and evaluators is not new. For instance, Bagues et al. [1] and Zinovyeva & Bagues [32] describe how these or similar connections relate to the potential candidates' decision to apply and their success in the research assessment exercise.

| ID | Description |
|---|---|
| cand_comm | number of citations going from a candidate's publication to a publication authored by at least one member of the commission |
| comm_cand | number of citations going from a publication authored by at least one member of the commission to a candidate's publication |
| BC (bibliographic coupling) | number of publications cited by both a publication authored by the candidate and a publication authored by at least one member of the commission |
| CC (co-citation) | number of publications citing both a publication authored by the candidate and a publication authored by at least one member of the commission |
| cand_other | number of other publications (i.e. which are not authored neither by the candidate nor by any member of the commission) cited by a publication authored by the candidate |
| other_cand | number of other publications (i.e. which are not authored neither by the candidate nor by any member of the commission) citing a publication authored by the candidate |

Table 2: **The metrics extracted from the citation networks with a short description**

The presence of a lot of citations from the candidates' to the commission members' publications, and vice versa, proves that they share research interests and activities. The same can be said about high values of co-citation and bibliographic coupling. These are all indicators of a strong relation between candidates and evaluators that might have played a role in the peer-review process of research assessment exercises.

Our interest is also to investigate the interaction between the metrics in Table 1 and Table 2. It might well happen that the citation network of a candidate has a few connections with that of the commission, but the candidate equally gets a positive evaluation in the research assessment exercise. In that case, the particular research interests of the candidate do not overlap with those of the commission but other metrics might have a higher impact.

To give readers a clearer idea of our approach, Figure 1 shows one example of the citation network we aim at building for each candidate. A graph is created in which the nodes represent publications and the edges represent citation links. Note that we do not use arrows for the direction of the citations, just to not overload the picture, but we take it into account when computing the metrics. Different colors are used to indicate if a publication is by the candidate (blue) or the commission member (red) or co-authored by the candidate and at least one member of the commission (green). The gray nodes indicate the publications of other authors - neither the candidate nor any member of the commission - that cite or are cited by a publication of the candidate or a member of the commission. These citations, in fact, contribute to the values of co-citations and bibliographic coupling.

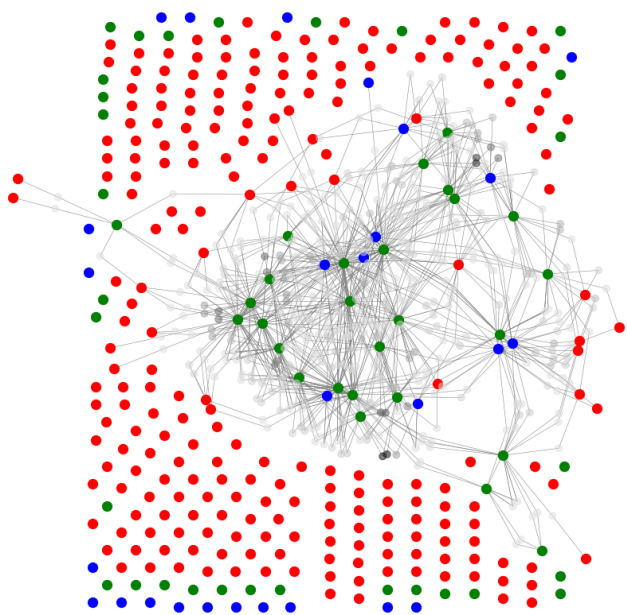

Figure 1: **Example of the kind of citation network we build for a candidate. In this example, the candidate has a strong co-authorship connection with the commission members (i.e. the green nodes) and a lot of candidate's and commission members' publications are closely-coupled by citation relationships, as confirmed by the dense network in the center of the diagram.**

## 2.2 Data

This methodology takes into consideration the bibliographic and citation data of the candidates and the commissions participating in any national scientific qualification that presents a human evaluation phase. It focuses on the publications of both the candidates applying to a research assessment exercise, and the commission members who are in charge of evaluating these candidates in the peer-review component of the exercise.

## 2.3 Sources

We aim at collecting the bibliographic and citation data of the candidates and commissions from four open access sources.

The first, Microsoft Academic Graph (https://www.microsoft.com/en-us/research/project/microsoft-academic-graph/)[29], referred to as MAG here, results from the efforts of the Microsoft Academic Search (MAS) project. This dataset is updated biweekly and is distributed under an open data license for research and commercial applications. We use a copy of MAG created and made available by Internet Archive in January 2020 [22].

The second, OpenAIRE Graph (https://www.openaire.eu/) [26], referred to as OA here, includes information about objects of the scholarly communication life-cycle (publications, research data, research software, projects, organizations, etc.) and semantic links among them. It is created bi-monthly, and is accessible for scholarly communication and research analytics. We use the dump that OpenAIRE has released on Zenodo in April 2021 [19].

The third, Crossref (https://www.crossref.org/)[15], referred to as CR here, was born as a nonprofit membership association among publishers to promote collaboration to speed research and innovation. The dataset is fully curated and governed by the members. We use the dump released in January 2021 [6].

The fourth, OpenCitations (https://opencitations.net/)[25], referred to as OC here, features COCI, the OpenCitations Index of Crossref Open DOI-to-DOI Citations [14]. COCI stores the citation links between citing and cited bibliographic entities identified by their DOIs, enabling the retrieval of their metadata from other sources. It is accessible through a variety of tools. We aim at using the freely available dump released in December 2020 [24].

## 2.4 Collection process

To collect the bibliographic and citation data of both the candidates and the commissions, we aim at using the following procedure.

First, we set up a database containing all the bibliographic and citation present in each dataset dump. We download and pre-process each dump. We take each publication, select the metadata of interest for our analysis (author, title, year, identifiers, references, citations) and save it to a smaller JSON file. Each JSON file is then imported as a collection into a MongoDB database. We then create a series of indexes to query publications by year and title, DOI and MAG identifiers.

Secondly, we search, in the database, for the bibliographic and citation data of the publications of the candidates and the commission members participating in the research assessment exercise. It is first necessary to have a list of all the publications of both the candidates and the commission members with the corresponding metadata (title, year, doi, authors etc.). Then, we search for each publication by querying the MAG collection either by doi, if present, or by year and title. For each found publication we collect the Paper Id, the Reference Ids and the Author Id related to the candidate or commission member of interest. The first identifies the entity, the second is a list of the Paper Ids of all cited entities, and the third identifies the author. Since in MAG each author can be assigned multiple Author Ids, we retrieve one for each publication and keep only unique ids. We then query MAG by each Author Id to retrieve all publications associated with the author. We do so to find publications in the list that are not found using the method described above, and to add any extra publication that is not present in the original list. The remaining publications that are not found in MAG and are not assigned a DOI, are searched for in OA by year and title; if still unfound, they are also searched for in CR.

In the fourth step, we retrieve the metadata of the citing and cited entities for each publication. In MAG, to collect the referenced entities metadata we query the database by Paper Id using each Reference Id of the publication; viceversa, to collect citing entities metadata, we query by Reference Id using the publication Paper Id. To ensure optimal coverage of citation data, for each publication associated with a DOI, we also query in COCI to collect the DOIs of the citing and cited publications. We then query these in CR to retrieve their metadata. Given that citation metadata from MAG and COCI might overlap, citing and cited entities are disambiguated by DOI to prevent repetitions.

Finally, candidate applications are divided into three sections, A, B or C, depending on how many of their originally listed publications are found in these open access datasets. Applications are placed in section A if more than 15 of the originally listed publications are retrieved, or, otherwise, 70% of them are retrieved. Applications are placed in section B when less than 70% of their original publications are found but additional publications are extracted from MAG to reach a total number of retrieved publications comparable to the original one. Applications are placed in section C otherwise. This categorization gives us the ability to decide whether to include applications that are not well covered in the datasets in the following step.

## 2.5 Machine Learning Classifiers

The last step of our methodology consists of using computational methods and machine learning techniques to investigate which metrics provide more insights on human evaluation procedures.

In particular, we employ two different machine learning classifiers: SVM and Random Forest. These classifiers have been recently used in many bibliometric studies, demonstrating their effectiveness in different contexts [7, 13, 27, 28].

We use the metrics extracted from bibliometric data and described in Section 2.2 as features of these classification algorithms for automatically discerning between academics who passed an evaluation procedure and who did not. This is a binary classification problem since we have two mutually exclusive classes (i.e. one composed of candidates who passed the evaluation procedure, and the other of candidates who failed) and, for each individual in the population, we attempt to predict which class he/she belongs to.

We base our analysis on a variable ranking method from feature selection theory [12] to determine the metrics that are significant for the current classification problem, and hence are useful to solve the classification problem and to identify the metrics that are either redundant or irrelevant – and that can, thus, be removed without incurring much loss of information [2].

We test each possible subset of metrics and find those which minimizes the error rate and lead to good predictions. We start by computing all the possible combinations of metrics that can be obtained considering the $m$ metrics in the dataset. The number of combinations can be computed by the following expression:

$$combinations_{total} = \sum_{k=1}^{m} C_{m,k} = \sum_{k=1}^{m} \frac{m!}{k! \, (m-k)!}$$

The two classifiers (i.e. SVM and Random Forest) are computed using all the metric combinations and relying on cross-fold validation instead of explicitly splitting the dataset in training and test sets. An oversampling technique with stratification can be applied to the minority class for managing imbalanced classes, i.e. in cases where the distribution of examples across the known classes is not equal [18]. In particular, we used the approach described in [17] that prescribes to resample the smaller class (by creating synthetic instances chosen at random from from the minority class) until it consisted of as many samples as the majority class.

Of all the computed classifiers, we consider only those with good discrimination abilities (i.e. those whose weighted average F1-score is at least 0.7, and which therefore lead to good classification performances). Since the classifiers were calculated using all possible combinations of metrics, each metric was used as a feature by half of the classifiers. We then count how many times each of the $m$ metrics has been used as a feature by the classifiers with good discrimination abilities, and identify as significant for the current classification problem those that have been used by more than 50% of these classifiers. We also identify the metrics that are not relevant or redundant as those that are used as features by a low fraction (i.e less than 35%) of the classifiers with high classification performances.

The results of this phase allow us to understand what metrics provide relevant information that can be used by machine learning algorithms to replicate the human decisions of an evaluation commission. The hypothesis here is that if a subset of metrics is used by most of the classifiers with good performances, it signals that such metrics are important proxies of the human evaluation process and outcome.

## 3   Discussion and conclusions

In order to assess our methodology, we are currently applying it to the Italian National Scientific Qualification (NSQ). The NSQ is an assessment process to evaluate researchers and academics. In Italy, in fact, it is mandatory to pass it in order to apply for academic permanent positions. The NSQ consists of two distinct qualification procedures designed to attest two different levels of scientific maturity of a scholar, as Full Professor (FP) and Associate Professor (AP).[1]

In the Italian regulations, the disciplines are organized in 14 different Scientific Areas (SAs), in turn grouped in 184 Recruitment Fields (RFs). For each RF, the Ministry of University and Research (MIUR) appoints an evaluation committee (a.k.a. a commission) composed of five full professors responsible for assessing applicants for associate and full professorships. In order to apply to the NSQ, candidates have to submit a curriculum vitae with detailed information about their research accomplishments. These CVs are the first input of our methodology, from which we extract the candidates' publications.

We tested a preliminary version of our methodology on the data of candidates and committee members that took part in the 2016, 2017, and 2018 terms of the NSQ in two disciplines (RFs): "Historical and General Linguistics" and "Mathematical Methods of Economics, Finance and Actuarial Sciences". The details of our initial experiment can be found in [3] and were very promising, involving 500 unique candidates and 10 unique commission members. The experiments on ML proved that citation-based metrics, such as *co-citation* and *bibliographic coupling*, are quite reliable proxies for the decision of the commission.

Starting from those results, we refined our methodology as presented here and we are now building the full experiment that will cover all disciplines and RFs.

One of the first questions we asked ourselves was about the availability of the data and their ability to support the ML techniques in the classification task. Table 4 reports the number of candidates (and the related CVs) we eventually deployed for each SA, in the training and test phases. The quite high number of candidates we plan to use and the amount of training and testing corpus makes us optimistic about the possibility of replicating the preliminary results for the full NSQ.

| Scientific area (SA) | FP | | AP | |
|---|---|---|---|---|
| | *train* | *test* | *train* | *test* |
| 01 - Mathematics and Informatics | 897 | 392 | 1205 | 584 |
| 02 - Physics | 832 | 434 | 1336 | 692 |
| 03 - Chemistry | 855 | 379 | 1498 | 532 |
| 04 - Earth Sciences | 290 | 164 | 448 | 273 |
| 05 - Biology | 1598 | 710 | 4317 | 2347 |
| 06 - Medicine | 2446 | 1021 | 4957 | 2302 |
| 07 - Agricultural and Veterinary Sciences | 567 | 406 | 1080 | 610 |
| 08 - Civil Engineering and Architecture | 571 | 425 | 1019 | 733 |
| 09 - Industrial & Information Engineering | 1323 | 779 | 2378 | 1177 |
| 10 - Antiquities, Philology, Literary Studies, Art History | 619 | 622 | 1182 | 1430 |
| 11 - History, Philosophy and Psychology | 773 | 534 | 1651 | 1313 |
| 12 - Law Studies | 394 | 444 | 788 | 703 |
| 13 - Economics and Statistics | 1026 | 746 | 1439 | 1060 |
| 14 - Political and Social Sciences | 322 | 235 | 801 | 705 |
| **TOTAL** | 12513 | 7291 | 24099 | 14461 |

Table 4: **Dataset for the classification experiment on the full NSQ (terms 2016, 2017 and 2018).**

One objection to our approach might be about the use of open data only instead of closed one. This might reduce the lists of publications and citations for some candidates or commission members, and thus under-estimate their production. This is undoubtedly true but does not hinder the effectiveness and replicability of our approach: we are only interested in open data "by choice" but extending the same approach to closed data is straightforward.

---

[1] It is worth mentioning that passing the NSQ does not grant a tenure position. Each University is responsible for creating new positions according to financial and administrative requirements.

A further threat-to-validity is related to the use of only two classifiers, SVM and Random Forest: we started from those considering their successful deployment in similar contexts but we are also working to include other ones in our pipeline. A further direction to extend our work is the application of the methodology to other research assessment procedures: this can be deployed with little effort if these procedures share the same structure of the NSQ (candidates submitting publications evaluated by committee members) and are built on top of data that can be collected from open citational datasets.